\documentclass[a4paper,onecolumn,11pt,accepted=2026-11-23]{quantumarticle}
\pdfoutput=1
\usepackage[utf8]{inputenc}
\usepackage[english]{babel}
\usepackage[T1]{fontenc}
\usepackage[numbers]{natbib}
\usepackage{amsmath}
\usepackage{hyperref}
\usepackage{amssymb}
\usepackage{mathtools}
\usepackage{graphicx}
\usepackage{xcolor}
\usepackage{booktabs}
\usepackage{tabularx}

\usepackage{listings}
\definecolor{lightergray}{gray}{0.9}
\DeclarePairedDelimiter\abs{\lvert}{\rvert}
\DeclarePairedDelimiter\set{\{}{\}}
\DeclarePairedDelimiter\parens{(}{)}
\DeclarePairedDelimiter\bracks{[}{]}
\DeclarePairedDelimiter\bra{\langle}{\rvert}
\DeclarePairedDelimiter\ket{\lvert}{\rangle}
\DeclarePairedDelimiterX\braket[2]{\langle}{\rangle}{#1 \delimsize\vert #2}
\DeclareMathOperator{\cre}{cre}

\newcommand{\calD}{\ensuremath{\mathcal{D}}}

\newcommand{\calU}{\ensuremath{\mathcal{U}}}

\newcommand{\ffsim}{\mbox{\textsf{ffsim}}}
\newcommand{\fqe}{\mbox{\textsf{FQE}}}
\newcommand{\pyscf}{\mbox{\textsf{PySCF}}}
\newcommand{\openfermion}{\mbox{\textsf{OpenFermion}}}
\newcommand{\qiskit}{\mbox{\textsf{Qiskit}}}
\newcommand{\numpy}{\mbox{\textsf{NumPy}}}
\newcommand{\scipy}{\mbox{\textsf{SciPy}}}
\newcommand{\aer}{\mbox{\textsf{Qiskit Aer}}}
\newcommand{\dppy}{\mbox{\textsf{DPPy}}}
\newcommand{\asv}{\mbox{\textsf{asv}}}

\begin{document}

\title{ffsim: faster simulation of fermionic quantum circuits}

\author{Kevin J. Sung}
\email{kevinsung@ibm.com}
\affiliation{IBM Quantum, IBM Research, Yorktown Heights, NY 10598, USA}

\author{Inho Choi}
\affiliation{College of Natural Sciences, KAIST, Daejeon 34141, Republic of Korea}

\author{Mirko Amico}
\affiliation{IBM Quantum, IBM Research, Yorktown Heights, NY 10598, USA}

\author{Bartholomew Andrews}
\affiliation{Institute for Theoretical Physics, ETH Zürich, 8093 Zürich, Switzerland}
\affiliation{Department of Physics, University of California, Berkeley, CA 94720, USA}

\author{Esra Ayantuna}
\affiliation{Department of Physics, Temple University, Philadelphia, PA 19122, USA}

\author{Hrishikesh Belagali}
\affiliation{Department of Computational Mathematics, Science, and Engineering, Michigan State University, East Lansing, MI 48824, USA}
\affiliation{Center for Quantum Computing, Science, and Engineering, Michigan State University, East Lansing, MI 48824, USA}

\author{Yukio Kawashima}
\affiliation{IBM Quantum, IBM Research, Tokyo 103-8510, Japan}

\author{Ryan LaRose}
\affiliation{Department of Computational Mathematics, Science, and Engineering, Michigan State University, East Lansing, MI 48824, USA}
\affiliation{Department of Electrical and Computer Engineering, Michigan State University, East Lansing, MI 48824, USA}
\affiliation{Department of Physics and Astronomy, Michigan State University, East Lansing, MI 48824, USA}
\affiliation{Center for Quantum Computing, Science, and Engineering, Michigan State University, East Lansing, MI 48824, USA}

\author{Wan-Hsuan Lin}
\affiliation{Computer Science Department, University of California, Los Angeles, CA 90094, USA}
\affiliation{IBM Quantum, IBM Research, Yorktown Heights, NY 10598, USA}

\author{David Omanovic}
\affiliation{Department of Physics, Norwegian University of Science and Technology, NO-7491 Trondheim, Norway}

\author{Samuele Piccinelli}
\affiliation{IBM Quantum, IBM Research, CH–8803 Rüschlikon, Switzerland}
\affiliation{Institute of Physics, EPFL, CH-1015 Lausanne, Switzerland}

\author{Javier Robledo Moreno}
\affiliation{IBM Quantum, IBM Research, Yorktown Heights, NY 10598, USA}

\author{Abdullah Ash Saki}
\affiliation{IBM Quantum, IBM Research, Cambridge, MA 02142, USA}

\author{James Shee}
\affiliation{Department of Chemistry, Rice University, Houston, TX 77005-1892, USA}
\affiliation{Department of Physics and Astronomy, Rice University, Houston, TX 77005-1892, USA}

\author{Soyoung Shin}
\affiliation{IBM Quantum, IBM Research, San Jose, CA 95141, USA}

\author{Sabyrzhan Tasbolatov}
\affiliation{Independent Researcher, Z05H0H6 Astana, Kazakhstan}

\author{Minh C. Tran}
\affiliation{IBM Quantum, IBM Research, Yorktown Heights, NY 10598, USA}

\author{Kento Ueda}
\affiliation{IBM Quantum, IBM Research, Cambridge, MA 02142, USA}

\author{Haimeng Zhang}
\affiliation{IBM Quantum, IBM Research, Yorktown Heights, NY 10598, USA}

\author{Mario Motta}
\affiliation{IBM Quantum, IBM Research, Yorktown Heights, NY 10598, USA}

\maketitle

\begin{abstract}
    We present \ffsim{}, an open-source software library for fast simulation of fermionic quantum circuits. \ffsim{} exploits conservation of particle number and the $z$~component of spin, symmetries present in a wide range of fermionic systems, to dramatically reduce memory usage and simulation time compared to general-purpose quantum circuit simulators. Compared to \fqe{}, a library with similar functionality, \ffsim{} differs in software design and is faster on a representative set of simulation benchmarks. Beyond state vector evolution by basic fermionic gates, \ffsim{} offers a number of additional features including variational ansatzes, Hamiltonian time evolution via Trotter-Suzuki product formulas, efficient sampling of Slater determinants, seamless integration with \qiskit{} and \pyscf{}, and comprehensive documentation. We demonstrate \ffsim{}'s capabilities on scientific applications, including a 64-qubit Hubbard model with 8 electrons. \ffsim{}'s source code and documentation are available via \mbox{\url{https://github.com/Qiskit/ffsim}}.
\end{abstract}

\newpage

\section{Introduction}

With the continuing development of quantum computing hardware and algorithms, the ability to test, validate, and characterize these algorithms through classical simulation becomes increasingly important. Classical simulation serves a dual role: it enables researchers to prototype algorithms and explore their behavior on tractable problem sizes, and it provides numerical benchmarks against which quantum hardware results can be compared. While many general-purpose quantum circuit simulators have been developed, specialized simulators that exploit problem structure can offer vastly greater performance within their domain. The simulation of fermionic systems, a highly anticipated application of quantum computers, is one such domain.

This paper describes \ffsim{}, a Python library for simulating fermionic quantum circuits that exploits conservation of particle number and the $z$~component of spin. These symmetries are present in many fermionic systems, from molecules and materials to idealized lattice models. The idea to exploit these symmetries for state vector simulation is not new, and has previously been implemented in the Fermionic Quantum Emulator (\fqe{})~\cite{rubin2021fqe}. While \ffsim{} and \fqe{} share a common purpose, they differ in software design, features, and performance. For example, in single-threaded benchmarks on an Apple M1 Max with 16 spatial orbitals at $1/2$ filling, \ffsim{} is $2.4\times$ faster than \fqe{} for quadratic Hamiltonian evolution and $8.4\times$ faster for molecular Hamiltonian action, with more dramatic speedups in other operations (see Section~\ref{sec:performance}). \ffsim{} integrates with \qiskit{}~\cite{qiskit} and \pyscf{}~\cite{pyscf-1,pyscf-2,pyscf-3}, widely used open-source software libraries for quantum computing and quantum chemistry. Through its integration with \qiskit{}, \ffsim{} can simulate arbitrary qubit circuits composed of Hamming weight-preserving gates, without reference to fermions. \ffsim{} also offers a number of additional features, including variational ansatzes, Hamiltonian time evolution, efficient sampling of Slater determinants, and comprehensive documentation. Table~\ref{tab:capabilities} summarizes \ffsim{}'s main capabilities and points to the sections describing them.

 As an exact state vector simulator, \ffsim{}'s resource usage scales exponentially with system size, but with a smaller prefactor compared to a general-purpose quantum circuit simulator. The smaller prefactor meaningfully extends the range of tractable systems, as we demonstrate by applying \ffsim{} to scientific applications that involve simulating quantum circuits of up to 64 qubits. While a general-purpose quantum circuit simulator would require 256 EiB (about $2.7\times10^{11}$ GiB) to store the state vector of 64 qubits using double precision, \ffsim{} uses only 19.3 GiB for the demonstrated system, a Hubbard model with 32 lattice sites and four electrons of each spin ($1/8$ filling) (see Section~\ref{sec:trotter_error}).

 The rest of this paper is organized as follows. Section~\ref{sec:background} introduces the mathematical background and notation used throughout the paper. Section~\ref{sec:methods} describes \ffsim{}'s state vector representation and the algorithms used to apply fermionic gates. Section~\ref{sec:software} describes the software design and core features of \ffsim{}. Section~\ref{sec:additional-features} presents additional features, including variational ansatzes, Hamiltonian time evolution, and efficient sampling of Slater determinants. Section~\ref{sec:comparison} compares \ffsim{} with other software libraries and presents performance benchmarks. Section~\ref{sec:applications} demonstrates scientific applications enabled by \ffsim{}. Section~\ref{sec:limitations} discusses limitations of the framework. Finally, Section~\ref{sec:conclusion} discusses future directions and concludes the paper.
 
\begin{table}[tbp]
    \centering
    \small
    \renewcommand{\arraystretch}{1.1}
    \begin{tabularx}{\textwidth}{@{}>{\raggedright\arraybackslash}p{0.23\textwidth} >{\raggedright\arraybackslash}X l@{}}
        \toprule
        \textbf{Capability} & \textbf{Supported functionality} & \textbf{Section} \\
        \midrule
        State vector simulation & Exact simulation within a sector of fixed particle number and $z$ component of spin; a universal set of symmetry-preserving fermionic gates. & \ref{sec:methods} \\
        \addlinespace
        Operator action and expectation values & Fermionic operator action through the \scipy{} \texttt{LinearOperator} interface, with specialized implementations for common Hamiltonians. & \ref{sec:operator-action} \\
        \addlinespace
        Variational ansatzes and optimization & Unitary cluster Jastrow (UCJ), local UCJ (LUCJ), and unitary coupled cluster singles and doubles (UCCSD); parameter initialization and optimization, including the linear method. & \ref{sec:variational-ansatzes} \\
        \addlinespace
        Double factorization & Factorization of molecular Hamiltonians and coupled-cluster $t_2$-amplitudes, including compressed factorizations optimized to use fewer terms. & \ref{sec:double-factorization} \\
        \addlinespace
        Hamiltonian time evolution & Exact evolution via \scipy{}; Trotter-Suzuki product formulas of arbitrary order for double factorized and diagonal Coulomb Hamiltonians. & \ref{sec:hamiltonian-time-evolution} \\
        \addlinespace
        Slater determinants & Construction, manipulation, and efficient sampling using a compact representation, without storing the full state vector. & \ref{sec:slater-sampling} \\
        \addlinespace
        General fermionic operators & Representation and algebraic manipulation of sums of products of creation and annihilation operators, including normal ordering. & \ref{sec:general-operators} \\
        \addlinespace
        \qiskit{} circuit simulation & Simulation of Hamming weight-preserving circuits after computational basis state preparation, with optional final measurements; \texttt{Sampler} and final-state-vector interfaces. & \ref{sec:qiskit} \\
        \addlinespace
        \qiskit{} circuit construction & Fermionic gates in the Jordan-Wigner representation, optimized gate decompositions, and a transpiler pass for orbital rotations and state preparation. & \ref{sec:qiskit} \\
        \addlinespace
        \pyscf{} interoperability & Compatible full configuration interaction (FCI) state vector addressing, allowing direct exchange of state vectors with \pyscf{}. & \ref{sec:state-vector-representation} \\
        \addlinespace
        Documentation & Explanatory pages, how-to guides, tutorials, and API references for all public functions and classes. & \ref{sec:documentation} \\
        \bottomrule
    \end{tabularx}
    \caption{Summary of the main capabilities of \ffsim{}. Section references provide further details and usage examples.}
    \label{tab:capabilities}
\end{table}

\section{Background and notation}
\label{sec:background}

\subsection{Fermionic creation and annihilation operators}

A fermionic system consists of a set of fermionic modes, each of which can be either empty or occupied by a fermion. A system of $N$ modes is described by annihilation operators $\set{a_p}_{p=0}^{N-1}$ satisfying the canonical anticommutation relations
\begin{align}
    a_p a_q + a_q a_p = 0, \qquad
    a_p a_q^\dagger + a_q^\dagger a_p = \delta_{pq}.
\end{align}
The adjoint $a_p^\dagger$ is called a creation operator. The number operator $n_p = a_p^\dagger a_p$ has eigenvalues 0 and 1, and all number operators share a mutual eigenvector of eigenvalue 0 denoted by $\ket{\text{vac}}$.

\subsection{Electronic configurations}
\label{sec:background-configurations}

The Hilbert space of $N$ fermionic modes is spanned by orthonormal basis vectors
\begin{align}
    \ket{I} = \cre(I)\ket{\text{vac}},
\end{align}
where $I \subseteq \set{0, \ldots, N-1}$ and $\cre(I)$ is an ordered product of creation operators on modes contained in $I$, with the modes sorted in ascending order from right to left. For example, $\ket{\set{0, 2, 3}} = a^\dagger_3 a^\dagger_2 a^\dagger_0 \ket{\text{vac}}$ and $\ket{\set{}} = \ket{\text{vac}}$. The basis vectors $\ket{I}$ are called electronic configurations, or simply configurations. We also use the same term to refer to the subsets $I$ themselves. Note that in this work, electronic configurations are Slater determinants, and not configuration state functions~\cite{helgaker2000molecular}.

In the context of quantum chemistry, the physical significance of the fermionic modes is often determined by a Hartree-Fock calculation~\cite{helgaker2000molecular}. In this case, the configuration $\ket{\set{0, \ldots, \eta - 1}}$, where $\eta$ is the number of electrons in the system, is the Hartree-Fock state, denoted $\ket{\text{Hartree-Fock}}$.

\subsection{Orbital rotations}
\label{sec:background-orbital-rotations}

An orbital rotation is a unitary operator $\calU$ such that
\begin{align}
    \calU a^\dagger_p \calU^\dagger = \sum_q U_{qp} a^\dagger_q
\end{align}
for some $N \times N$ unitary matrix $U$. In an abuse of notation, we write $\calU(U)$ for the operator corresponding to $U$. It is a theorem (see Section 3.2 of Ref.~\cite{helgaker2000molecular} for a proof) that
\begin{align}
    \calU(U) = \exp\left(\sum_{pq} \log(U)_{pq} a^\dagger_p a_q\right).
\end{align}
The map $U \mapsto \calU(U)$ is a group homomorphism, that is,
\begin{align}
     \calU(UV) = \calU(U) \calU(V) \qquad \text{and} \qquad \calU(U^\dagger) = \calU(U)^\dagger
\end{align}
for any pair of unitary matrices $U$ and $V$.

\subsection{Slater determinants}
\label{sec:background-slater}

A Slater determinant is a state of the form
\begin{align}
    \ket{\text{Slater}} = \calU \ket{I}
\end{align}
where $\calU$ is an orbital rotation and $\ket{I}$ is an electronic configuration. Thus, electronic configurations are Slater determinants where the orbital rotation is the identity operator.

\subsection{Spin}
\label{sec:background-spin}

In the systems we consider, the fermionic modes come in pairs called spatial orbitals. Each spatial orbital $p \in \set{0, \ldots, N-1}$ contains two fermionic modes called spin orbitals, labeled by spin $\sigma \in \set{\alpha, \beta}$ (spin-up and spin-down). We write $a_{p\sigma}$ and $n_{p\sigma} = a_{p\sigma}^\dagger a_{p\sigma}$ for the corresponding annihilation and number operators. In systems with spin, we use $N$ to denote the number of spatial orbitals, so the total number of spin orbitals is $2N$. The electronic configurations take the form $\ket{I_\alpha, I_\beta} = \cre_\beta(I_\beta) \cre_\alpha(I_\alpha) \ket{\text{vac}}$, where $I_\alpha$ is a subset of $\set{0, \ldots, N-1}$ with $N_\alpha$ elements, $I_\beta$ is a subset of $\set{0, \ldots, N-1}$ with $N_\beta$ elements, and $\cre_\sigma(I)$ is an ordered product of creation operators with spin $\sigma$.

\subsection{Symmetry}

Two physically relevant symmetries are the total particle number $\mathcal{N} = \sum_{p\sigma} n_{p\sigma}$ and projected spin $\mathcal{S}_z = \frac{1}{2}\sum_p (n_{p\alpha} - n_{p\beta})$. When both are conserved, the Hilbert space decomposes into sectors labeled by $(N_\alpha, N_\beta)$, where $N_\sigma$ is the number of spin-$\sigma$ electrons. Each sector has dimension $\binom{N}{N_\alpha} \times \binom{N}{N_\beta}$, and \ffsim{} restricts simulation to a single such sector.

\subsection{Hamiltonian representations}

\ffsim{} works with the various Hamiltonian representations described in this section.

\subsubsection{Molecular Hamiltonians}

A molecular Hamiltonian has the form
\begin{align}
    \label{eq:molecular-hamiltonian}
    H = \sum_{\substack{pq \\ \sigma}} h_{pq} a^\dagger_{p\sigma} a_{q\sigma}
    + \frac12 \sum_{\substack{pqrs \\ \sigma \tau}} h_{pqrs}
    a^\dagger_{p\sigma} a^\dagger_{r\tau} a_{s\tau} a_{q\sigma},
\end{align}
where $h_{pq}$ and $h_{pqrs}$ are complex-valued tensors.

\subsubsection{Double factorized Hamiltonians}

A double factorized Hamiltonian~\cite{motta2021lowrank} has the form
\begin{align}
    \label{eq:double-factorized-hamiltonian}
    H = \sum_{\substack{pq \\ \sigma}} h_{pq} a^\dagger_{p\sigma} a_{q\sigma}
    + \frac12 \sum_t \sum_{\substack{pq \\ \sigma\tau}}
    J^{(t)}_{pq} n^{(t)}_{p\sigma} n^{(t)}_{q\tau},
\end{align}
where $n^{(t)}_{p\sigma} = \sum_{ij} U^{(t)}_{ip} a^\dagger_{i\sigma} a_{j\sigma} U^{(t)*}_{jp}$ is the number operator in the rotated basis defined by the unitary matrix $U^{(t)}$, and each $J^{(t)}$ is a real symmetric matrix.

\subsubsection{Diagonal Coulomb Hamiltonians}

A diagonal Coulomb Hamiltonian~\cite{babbush2018lowdepth,kivlichan2018quantumsimulation} has the form
\begin{align}
    \label{eq:diagonal-coulomb-hamiltonian}
    H = \sum_{\substack{pq \\ \sigma}} h_{pq} a^\dagger_{p\sigma} a_{q\sigma}
    + \frac12 \sum_{\substack{pq \\ \sigma \tau}} J^{\sigma \tau}_{pq} n_{p\sigma} n_{q\tau},
\end{align}
where each $J^{\sigma \tau}$ is a real symmetric matrix.

\section{State vector simulation}
\label{sec:methods}

\begin{table}[tbp]
    \centering
    \small
    \renewcommand{\arraystretch}{1.1}
    \begin{tabularx}{\textwidth}{@{}l@{\hspace{3em}}>{\raggedright\arraybackslash}X@{}}
        \toprule
        \textbf{Name} & \textbf{Operation} \\
        \midrule
        Number interaction & $\exp\left(i \theta n_{p\sigma}\right)$ \\
        \addlinespace
        Number-number interaction & $\exp\left(i \theta n_{p\sigma} n_{q\tau}\right)$ \\
        \addlinespace
        Givens rotation & $\exp\left(\theta (a^\dagger_{p\sigma} a_{q\sigma} - a^\dagger_{q\sigma} a_{p\sigma})\right)$ \\
        \addlinespace
        Tunneling interaction & $\exp\left(i \theta (a^\dagger_{p\sigma} a_{q\sigma} + a^\dagger_{q\sigma} a_{p\sigma})\right)$ \\
        \addlinespace
        Number operator sum evolution & $\exp\left(-i t \sum_{p\sigma} \lambda^{(\sigma)}_p n_{p\sigma}\right)$ \\
        \addlinespace
        Diagonal Coulomb evolution & $\exp\left(-\frac12 i t \sum_{pq\sigma\tau} J^{(\sigma \tau)}_{pq} n_{p\sigma} n_{q\tau}\right)$ \\
        \addlinespace
        Orbital rotation & $\exp\left(\sum_{pq\sigma} \log(U^{(\sigma)})_{pq} a^\dagger_{p\sigma} a_{q\sigma}\right)$ \\
        \addlinespace
        Quadratic Hamiltonian evolution & $\exp\left(-i t \sum_{pq\sigma} M^{(\sigma)}_{pq} a^\dagger_{p\sigma} a_{q\sigma}\right)$ \\
        \bottomrule
    \end{tabularx}
    \caption{
        A universal set of fermionic gates implemented in \ffsim{}. Here, $p, q \in \set{0, \ldots, N-1}$ index spatial orbitals and $\sigma, \tau \in \set{\alpha, \beta}$ are spin indices. $\theta$ is a real number, each $\lambda^{(\sigma)}$ is a real-valued vector, each $J^{(\sigma \tau)}$ is a real-valued matrix, each $U^{(\sigma)}$ is a unitary matrix, and each $M^{(\sigma)}$ is a Hermitian matrix.
    }
    \label{tab:gate-functions}
\end{table}

The core function of \ffsim{} is to store a fermionic wave function with a fixed particle number and $z$~component of spin, and evolve it by the application of fermionic quantum gates. Table~\ref{tab:gate-functions} lists some of the gates implemented in \ffsim{}. These gates form a universal set in the sense that any quantum circuit that conserves particle number and the $z$~component of spin can be decomposed into a sequence of these gates~\cite{bravyi2002fermionic}. There is some redundancy in the listed gates:
\begin{itemize}
    \item The number interaction is a special case of number operator sum evolution.
    \item The number-number interaction is a special case of diagonal Coulomb evolution.
    \item The tunneling interaction can be implemented using a Givens rotation and number interactions.
    \item The Givens rotation is a special case of orbital rotation.
    \item Quadratic Hamiltonian evolution with matrix $M$ and time $t$ is equivalent to an orbital rotation with $U = \exp(-i M t)$.
\end{itemize}
The rest of this section explains how we represent a fermionic wave function and evolve it by the application of fermionic gates.

\subsection{State vector representation}
\label{sec:state-vector-representation}

The fermionic wave function is represented as
\begin{align}
    \ket{\Psi} = \sum_{I_\alpha, I_\beta} \gamma(I_\alpha, I_\beta) \ket{I_\alpha, I_\beta},
\end{align}
where $\gamma(I_\alpha, I_\beta)$ is the coefficient of the electronic configuration $\ket{I_\alpha, I_\beta}$. As noted in Section~\ref{sec:background-configurations}, the configurations here are Slater determinants, and not configuration state functions. This state vector representation is commonly used in software for full configuration interaction (FCI)~\cite{knowles1984new,olsen1988determinant}. \ffsim{} uses \pyscf{}'s~\cite{pyscf-1,pyscf-2,pyscf-3} FCI module to assign state vector addresses to the basis vectors $\ket{I_\alpha, I_\beta}$.

The dimension of the state vector is $D = d_\alpha d_\beta$, where $d_\sigma = \binom{N}{N_\sigma}$, $N$ is the number of spatial orbitals, $N_\alpha$ the number of spin-up electrons, and $N_\beta$ the number of spin-down electrons. In contrast, a general-purpose quantum circuit simulator would store a state vector of dimension $2^{2N}$, since there are $2N$ spin orbitals and each spin orbital gets mapped to one qubit under standard fermionic mappings such as the Jordan-Wigner transformation~\cite{jordan1928uber,bravyi2002fermionic}.

\subsection{Fermionic gate application}

To evolve the state vector by the application of a fermionic gate, we need to compute the new coefficients $\gamma'(I_\alpha, I_\beta)$ of the resulting state. This section explains how this works for the number operator sum evolution, diagonal Coulomb evolution, and orbital rotation. All other fermionic gates are special cases of these or can be implemented by composing them.

All state vector complexity estimates in this paper assume fixed filling, with both spin filling fractions $N_\alpha/N$ and $N_\beta/N$ held constant and strictly between zero and one. As a consequence, both $d_\alpha$ and $d_\beta$ grow exponentially with $N$.

\subsubsection{Number operator sum evolution}

Number operator sum evolution takes as input a real-valued vector $\lambda$ of length $N$ and a real number $t$ representing the evolution time, and applies the operator
\begin{align}
    \exp\left(-i t \sum_{p\sigma} \lambda_p n_{p\sigma}\right).
\end{align}
The new coefficients are equal to the original coefficients multiplied by a phase:
\begin{align}
    \gamma'(I_\alpha, I_\beta) = \exp\parens*{-i t \bracks*{\sum_{p \in I_\alpha} \lambda_p + \sum_{p \in I_\beta} \lambda_p}} \gamma(I_\alpha, I_\beta).
\end{align}
\ffsim{} supports spin-dependent interactions by allowing the user to pass two separate vectors $\lambda^{(\alpha)}$ and $\lambda^{(\beta)}$ specifying different coefficients for each spin.

The phase separates into spin-up and spin-down factors, each of which is computed once per configuration of that spin and reused. Computing these factors costs $O(N(d_\alpha + d_\beta))$, and updating the state vector amplitudes costs $O(D)$. Since both $d_\alpha$ and $d_\beta$ grow exponentially with $N$ and $D = d_\alpha d_\beta$, the amplitude updates dominate, giving a total cost of $O(D)$. The amplitude updates are performed in place and can be distributed across threads.

\subsubsection{Diagonal Coulomb evolution}

Diagonal Coulomb evolution takes as input an $N \times N$ real symmetric matrix $J$ and a real number $t$ representing the evolution time, and applies the operator
\begin{align}
    \exp\left(-\frac12 i t \sum_{pq\sigma\tau} J_{pq} n_{p\sigma} n_{q\tau}\right).
\end{align}
The new coefficients are equal to the original coefficients multiplied by a phase:
\begin{align}
    \gamma'(I_\alpha, I_\beta) = \exp\parens*{-\frac12 i t \sum_{(p, q) \in S} J_{pq}} \gamma(I_\alpha, I_\beta).
\end{align}
Here, the indices in the summation range over a multiset $S$ that contains pairs $(p, q)$ such that either
\begin{itemize}
    \item $p \in I_\alpha$ and $q \in I_\alpha$,
    \item $p \in I_\alpha$ and $q \in I_\beta$,
    \item $p \in I_\beta$ and $q \in I_\alpha$, or
    \item $p \in I_\beta$ and $q \in I_\beta$.
\end{itemize}
\ffsim{} supports spin-dependent interactions by allowing the user to pass three separate matrices $J^{(\alpha \alpha)}, J^{(\alpha \beta)}$, and $J^{(\beta \beta)}$ specifying different coefficients for different combinations of spins.

\ffsim{} precomputes the same-spin phases and, for each spin-up configuration, its interaction phase with each spin-down orbital. This costs $O(N^2(d_\alpha + d_\beta))$. Combining these factors requires $O(N)$ work per amplitude, giving a total cost of $O(N^2(d_\alpha + d_\beta) + ND) = O(ND)$. Both the phase computation and the in-place amplitude updates can be distributed across threads.

\subsubsection{Orbital rotation}
\label{sec:orbital-rotation}

Orbital rotation takes as input an $N \times N$ unitary matrix $U$, and applies the operator
\begin{align}
\exp\left(\sum_{pq\sigma} \log(U)_{pq} a^\dagger_{p\sigma} a_{q\sigma}\right).
\end{align}
We implement an orbital rotation by decomposing it into a sequence of elementary operations called Givens rotations. This method was described in Ref.~\cite{atchity1999orbitaltransformations} using the term \emph{Jacobi rotations}, and rediscovered in the context of quantum computing in Ref.~\cite{wecker2015solvingstrongly} using the term \emph{Givens rotations}, which we adopt. A Givens rotation is described by a real number $c$ and a complex number $s$ such that the matrix
\begin{align}
  \begin{pmatrix}
    c & s \\
    -s^* & c
  \end{pmatrix}
\end{align}
is unitary, and it can be applied to two orbitals $p$ and $q$ with the same spin $\sigma$. When $\sigma = \alpha$, its effect on the state vector coefficients is given by
\begin{align}
    \gamma'(I_\alpha, I_\beta) =
    \begin{cases}
        c \gamma(I_\alpha, I_\beta) + s \gamma(I_\alpha^{p \mapsto q}, I_\beta) & \text{if $p \in I_\alpha$ and $q \notin I_\alpha$}, \\
        -s^* \gamma(I_\alpha^{q \mapsto p}, I_\beta) + c \gamma(I_\alpha, I_\beta) & \text{if $q \in I_\alpha$ and $p \notin I_\alpha$}, \\
        \gamma(I_\alpha, I_\beta) & \text{otherwise},
    \end{cases}
\end{align}
and similarly when $\sigma = \beta$. Here, the notation $I_\sigma^{p \mapsto q}$ denotes the set obtained from $I_\sigma$ by replacing $p$ with $q$.
The action of a Givens rotation on the state vector can be implemented using the \texttt{zrot} function from \textsf{LAPACK}~\cite{lapack}, the standard linear algebra software library.
\ffsim{} supports spin-dependent interactions by allowing the user to pass two separate matrices $U^{(\alpha)}$ and $U^{(\beta)}$ specifying different orbital rotations for each spin.

Decomposing a dense $N \times N$ unitary matrix takes $O(N^3)$ operations and produces $O(N^2)$ Givens rotations and $O(N)$ single-orbital phase shifts per spin. Each rotation or phase shift costs $O(D)$, so applying the full sequence costs $O(N^2 D)$. With configuration-address maps precomputed, the total cost is therefore $O(N^3 + N^2 D)$, dominated by state vector updates for large $D$. In \ffsim{}, these updates use a dedicated Rust kernel that modifies pairs of amplitudes in place, with independent pairs distributed across threads. The address maps are cached and reused across orbital rotations.

\section{Software design and core features}
\label{sec:software}

\ffsim{} aims for a simple design, using established interfaces where possible. While \ffsim{}'s public interface is in Python, some features are implemented internally using Rust for improved performance. These Rust extensions are compiled to binaries and distributed with \ffsim{} on all supported platforms.

\subsection{State vector representation and manipulation}

A state vector is represented as a one-dimensional \numpy{}~\cite{harris2020numpy} array. In addition to the array of coefficients, manipulation of the state vector requires knowledge of the number of spatial orbitals and the numbers of spin-up and spin-down electrons. Functions that manipulate the state vector accept these values as additional arguments. For example, the following code applies an orbital rotation to a state vector:
\begin{lstlisting}
rotated_vec = ffsim.apply_orbital_rotation(vec, orbital_rotation, norb=norb, nelec=nelec)
\end{lstlisting}
Here, \texttt{orbital\_rotation} is the $N \times N$ unitary matrix describing the orbital rotation, \texttt{norb} is an integer giving the number of spatial orbitals, and \texttt{nelec} is a tuple of two integers giving the numbers of spin-up and spin-down electrons. This example demonstrates the basic usage pattern of \ffsim{}, where fermionic gate application is performed by functions that accept the initial state vector as input and return the evolved state vector.

\subsection{Operator action and computing expectation values}
\label{sec:operator-action}

In addition to evolving a state vector by the unitary action of a fermionic gate, \ffsim{} also supports computing the generally non-unitary action of a fermionic operator, such as a Hamiltonian, on the state vector. This feature is necessary, for example, to compute expectation values. Operator action is implemented via the \texttt{LinearOperator} interface of \scipy{}~\cite{scipy}. Operator action for arbitrary fermionic operators is supported, with specialized implementations available for common operators, such as the molecular Hamiltonian~(\ref{eq:molecular-hamiltonian}). This Hamiltonian is represented by the \texttt{MolecularHamiltonian} class, which can be initialized as follows:
\begin{lstlisting}
mol_ham = ffsim.MolecularHamiltonian(one_body_tensor, two_body_tensor)
\end{lstlisting}
Here, \texttt{one\_body\_tensor} is a matrix storing the coefficients $h_{pq}$, and \texttt{two\_body\_tensor} is a tensor storing the coefficients $h_{pqrs}$.
\ffsim{} provides a protocol for objects that can be converted to \scipy{} \texttt{LinearOperator}s, with the \texttt{linear\_operator} function being used to perform the conversion. For example, the following code obtains the \texttt{LinearOperator} for the molecular Hamiltonian:
\begin{lstlisting}
linop = ffsim.linear_operator(mol_ham, norb=norb, nelec=nelec)
\end{lstlisting}
The code to compute an expectation value $\bra{\Psi} H \ket{\Psi}$ then follows standard \numpy{} syntax:
\begin{lstlisting}
energy = np.vdot(vec, linop @ vec)
\end{lstlisting}
Conveniently, \texttt{LinearOperator}s can be passed to \scipy{}'s sparse linear algebra routines. For example, you can compute the exact time evolution by the Hamiltonian $e^{-i H t} \ket{\Psi}$ using the \texttt{scipy.sparse.linalg.expm\_multiply} function:
\begin{lstlisting}
evolved_vec = scipy.sparse.linalg.expm_multiply(-1j * time * linop, vec)
\end{lstlisting}

\begin{figure}
    \centering
    \includegraphics[width=\linewidth]{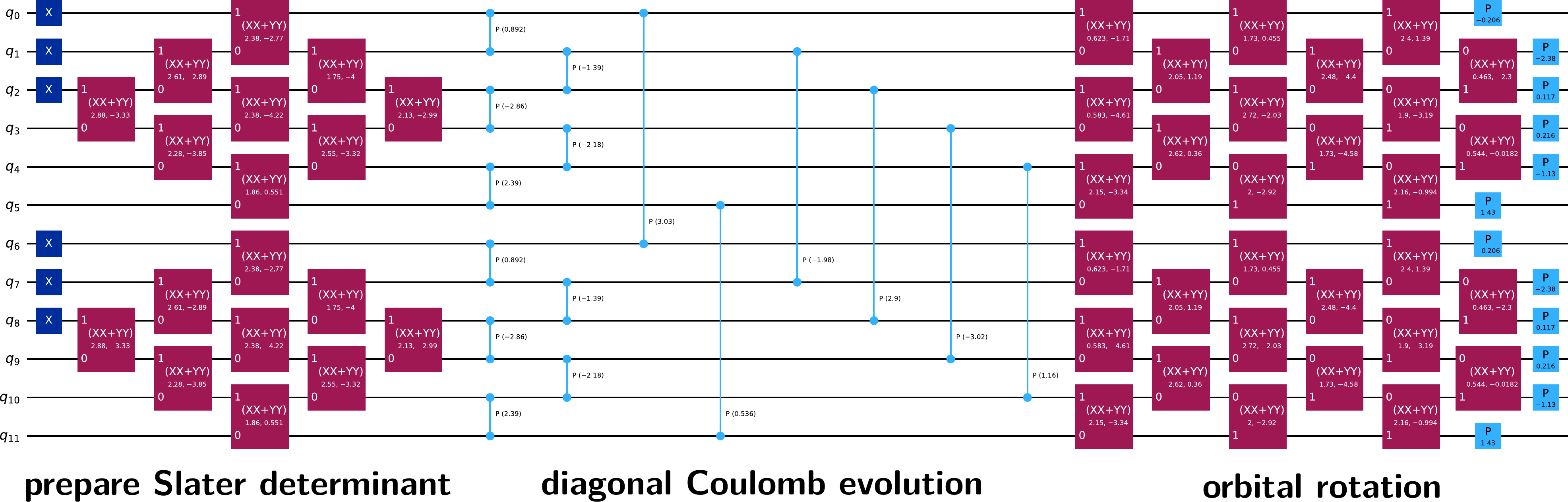}
    \caption{A \qiskit{} circuit that \ffsim{} can simulate. This circuit implements an instance of the local unitary Cluster Jastrow (LUCJ) ansatz. First, a Slater determinant is prepared by applying a diamond-shaped pattern of Givens rotations (as \texttt{XXPlusYYGate}s) to a computational basis state. Then, a diagonal Coulomb evolution is implemented using controlled-phase gates. Finally, an orbital rotation is implemented by a brickwork pattern of Givens rotations followed by a layer of single-qubit phase gates.}
    \label{fig:qiskit-lucj-circuit}
\end{figure}

\subsection{Integration with \qiskit{}}
\label{sec:qiskit}

Besides the basic interface where you manipulate the state vector directly using functions, \ffsim{} also supports simulating quantum circuits constructed using \qiskit{}~\cite{qiskit}, a widely used Python library for quantum computing. Note that the \qiskit{} circuit need not be related to fermionic simulations; the only requirement is that it is composed of gates that preserve Hamming weight (such gates include \texttt{CPhaseGate}, \texttt{SwapGate}, and \texttt{XXPlusYYGate}, among others). Specifically, \ffsim{} can simulate \qiskit{} circuits that have the following structure:
\begin{enumerate}
    \item The circuit begins by preparing a computational basis state, with a series of \texttt{X} gates applied to the all zeros state.
    \item Next, the circuit contains a sequence of unitary gates that preserve Hamming weight.
    \item Finally, any measurements occur only at the end of the circuit.
\end{enumerate}
To support simulation of \qiskit{} circuits, \ffsim{} provides an implementation of the \qiskit{} \texttt{Sampler} primitive, as well as a function that takes a \qiskit{} circuit as input and returns the final state vector.

In addition to simulating gates defined in \qiskit{} itself, \ffsim{} also defines new \qiskit{} gates to represent fermionic gates in the Jordan-Wigner representation~\cite{jordan1928uber,bravyi2002fermionic}, and implements highly optimized decompositions into lower-level gates. It includes a \qiskit{} transpiler pass that merges orbital rotations and detects when they are applied to computational states, in which case they can be replaced by a state preparation gate with a more efficient gate decomposition~\cite{jiang2018quantumalgorithms}. Thus, \ffsim{} can produce highly optimized \qiskit{} circuits for fermionic simulations, ready to be executed on quantum computers.

Figure~\ref{fig:qiskit-lucj-circuit} shows an example of an optimized \qiskit{} circuit for implementing a local unitary cluster Jastrow (LUCJ) ansatz~\cite{motta2023bridgingphysical} with a single repetition applied to the Hartree-Fock state, followed by an orbital rotation. A single repetition of the LUCJ ansatz has the form $\calU \calD \calU^\dagger$ where $\calU$ is an orbital rotation and $\calD$ is a diagonal Coulomb evolution, so the circuit prepares a state of the form
\begin{align}
    \calU_1 \calU_0 \calD \calU_0^\dagger \ket{\text{Hartree-Fock}}
\end{align}
where $\calU_0$ and $\calU_1$ are orbital rotations. Since the product of orbital rotations is another orbital rotation (see Section~\ref{sec:background-orbital-rotations}), and an orbital rotation applied to the Hartree-Fock state yields a Slater determinant (see Section~\ref{sec:background-slater}), the state can be rewritten as
\begin{align}
    \calU \calD \ket{\text{Slater}}
\end{align}
where $\calU = \calU_1 \calU_0$ is an orbital rotation and $\ket{\text{Slater}}$ is a Slater determinant. \ffsim{} can automatically perform this simplification. As a result, the final circuit prepares a Slater determinant using an optimized orbital rotation decomposition, then applies a diagonal Coulomb evolution, and ends with a general orbital rotation. The gate decompositions of each operation are shown in the figure, with Slater determinant preparation and orbital rotations being implemented with Givens rotations~\cite{jiang2018quantumalgorithms,clements2016optimal}, and diagonal Coulomb evolution with controlled-phase gates.

\subsection{Comprehensive documentation}
\label{sec:documentation}

\ffsim{} strives to uphold high standards for documentation, offering explanatory pages, how-to guides, tutorials, and detailed API references for all publicly exposed functions and classes. The documentation website is located at \url{https://qiskit.github.io/ffsim/}.

\section{Additional features}
\label{sec:additional-features}

In addition to the core features described above, \ffsim{} includes a number of additional features, some of which we describe in this section.

\subsection{Variational ansatzes}
\label{sec:variational-ansatzes}

Variational ansatzes are parameterized quantum circuits used by quantum algorithms, including the variational quantum eigensolver (VQE)~\cite{peruzzo2014vqe} as well as sample-based algorithms like quantum-selected configuration interaction (QSCI)~\cite{kanno2023qsci}, that require heuristic state preparation or evolution. \ffsim{} includes implementations of the unitary cluster Jastrow (UCJ) ansatz~\cite{matsuzawa2020jastrowtype} and its variant known as local UCJ (LUCJ)~\cite{motta2023bridgingphysical}, as well as the unitary coupled cluster, singles and doubles (UCCSD) ansatz~\cite{bartlett1989alternative,peruzzo2014vqe}. Users can also readily implement their own ansatzes as long as they conserve particle number and the $z$~component of spin. Ansatz classes store the parameters specifying their unitary operation, and can be applied to a state vector using the \texttt{ffsim.apply\_unitary} Python protocol function.

As an example, the following code initializes a real-valued restricted UCCSD operator from $t_1$- and $t_2$-amplitude tensors, and then applies the ansatz operator to a state vector:
\begin{lstlisting}
uccsd_op = ffsim.UCCSDOpRestrictedReal(t1, t2)
vec = ffsim.apply_unitary(vec, uccsd_op, norb=norb, nelec=nelec)
\end{lstlisting}
The UCJ and LUCJ ansatzes can be initialized from $t_2$-amplitudes via double factorization~\cite{matsuzawa2020jastrowtype,motta2023bridgingphysical}, with $t_1$-amplitudes optionally being used to initialize a final orbital rotation~\cite{robledomoreno2023enhancing,lin2025improved}:
\begin{lstlisting}
ucj_op = ffsim.UCJOpSpinBalanced.from_t_amplitudes(t2, t1=t1)
\end{lstlisting}
To facilitate variational optimization using standard numerical optimizers (such as those implemented in \scipy{}), ansatz classes provide methods to initialize from and convert to a flat vector of real-valued parameters. \ffsim{} also includes an implementation of the linear method~\cite{toulouse2007optimization,motta2015implementation,motta2024quantum} for variational optimization, a gradient-based algorithm that can outperform common alternatives such as L-BFGS-B~\cite{byrd1995lbfgsb,zhu1997algorithm}. See Ref.~\cite{motta2024quantum} for a comparison of the linear method and L-BFGS-B applied to VQE, with simulation data collected using \ffsim{}.

\subsection{Compressed double factorization}
\label{sec:double-factorization}

The double factorized representation~(\ref{eq:double-factorized-hamiltonian}) of the molecular Hamiltonian can significantly reduce the cost of performing time evolution and measurement on a quantum computer, compared to the standard representation~(\ref{eq:molecular-hamiltonian})~\cite{motta2021lowrank,huggins2021efficient}. Besides the molecular Hamiltonian, double factorization can also be applied to $t_2$-amplitudes from CCSD or UCCSD, which is useful for implementing Trotterized UCCSD and for producing physically motivated initial parameters for the UCJ and LUCJ ansatzes~\cite{motta2021lowrank,lin2025improved}. \ffsim{} includes functions for performing double factorization of both molecular Hamiltonians and $t_2$-amplitudes. In both cases, \ffsim{} supports performing a compressed double factorization, which optimizes the factorization to approximate the original tensor using fewer terms~\cite{cohn2021quantumfilter,lin2025improved}.

The UCJ example in the previous section already showed the high-level interface for initializing a UCJ operator via double factorization of $t_2$-amplitudes, and the compressed factorization can be enabled by setting the desired number of ansatz repetitions and enabling the \texttt{optimize} flag:
\begin{lstlisting}
ucj_op = ffsim.UCJOpSpinBalanced.from_t_amplitudes(t2, t1=t1, n_reps=3, optimize=True)
\end{lstlisting}
The syntax to initialize a \texttt{DoubleFactorizedHamiltonian} from a \texttt{MolecularHamiltonian} is similar:
\begin{lstlisting}
df_ham = ffsim.DoubleFactorizedHamiltonian.from_molecular_hamiltonian(
    mol_ham, max_vecs=10, optimize=True
)
\end{lstlisting}
\ffsim{} also exposes lower-level linear algebra routines that accept and return \numpy{} arrays directly:
\begin{lstlisting}
diag_coulomb_mats, orbital_rotations = ffsim.linalg.double_factorized_t2(t2)
diag_coulomb_mats, orbital_rotations = ffsim.linalg.double_factorized(two_body_tensor)
\end{lstlisting}

\subsection{Hamiltonian time evolution}
\label{sec:hamiltonian-time-evolution}

Hamiltonian time evolution is a fundamental algorithmic primitive for quantum computation~\cite{lloyd1996universal,childs2010ontherelationship,low2017optimal}. As mentioned in Section~\ref{sec:operator-action}, exact time evolution can be performed on a \scipy{} \texttt{LinearOperator} representation of the Hamiltonian using \scipy{}'s \texttt{expm\_multiply} function. \ffsim{} also supports approximate Hamiltonian time evolution using Trotter-Suzuki product formulas, a common approach to Hamiltonian simulation on quantum computers. Trotter simulation applies to a Hamiltonian expressed as a sum of terms, $H = \sum_{k=1}^L H_k$, where it is assumed that time evolution under individual terms $H_k$ can be performed exactly. The total evolution time $T$ is first divided into a number of smaller time steps called Trotter steps: $\exp(-i H T) = \exp(-i H t)^r$ where $t = T / r$ and $r$ is the number of Trotter steps. Then, time evolution for a single Trotter step $\exp(-i H t)$ is approximated by a product formula. \ffsim{} supports product formulas of arbitrary order as defined in Ref.~\cite{suzuki1991general}:
\begin{align*}
    S_0(t) &= \exp\parens*{-i H_L t} \cdots \exp\parens*{-i H_1 t}\\
    S_1(t) &= \exp\parens*{-i H_1 t/2}\cdots\exp\parens*{-i H_L t/2}\exp\parens*{-i H_L t/2}\cdots\exp\parens*{-i H_1 t/2}\\
    S_k(t) &= S_{k-1}^2(u_k t) S_{k-1}\parens*{(1-4u_k)t} S_{k-1}^2(u_k t)
\end{align*}
where $u_k = 1/(4-4^{1/(2k-1)})$. Note that for programming convenience, the order in \ffsim{} follows the sequence $0, 1, 2, 3, \ldots$, which differs from the usual $1, 2, 4, 6, \ldots$ found in the existing literature~\cite{childs2018toward,childs2021theory}.

\ffsim{} includes functions for Trotter simulation of the double factorized Hamiltonian~(\ref{eq:double-factorized-hamiltonian}) and diagonal Coulomb Hamiltonian~(\ref{eq:diagonal-coulomb-hamiltonian}). For both Hamiltonians, the one-body term $\sum_{pq\sigma} h_{pq} a^\dagger_{p\sigma} a_{q\sigma}$ can be implemented exactly by an orbital rotation. For the diagonal Coulomb Hamiltonian, the two-body term $\frac12 \sum_{pq\sigma \tau} J^{\sigma \tau}_{pq} n_{p\sigma} n_{q\tau}$ can be implemented exactly via diagonal Coulomb evolution, and for the double factorized Hamiltonian, each $\sum_{ij\sigma\tau} J^{(t)}_{ij} n^{(t)}_{i\sigma} n^{(t)}_{j\tau}$ term can be implemented exactly by a combination of orbital rotation and diagonal Coulomb evolution.

The following code applies Trotterized time evolution by a double factorized Hamiltonian:
\begin{lstlisting}
evolved_vec = ffsim.simulate_trotter_double_factorized(
    vec,
    df_hamiltonian,
    time,
    norb=norb,
    nelec=nelec,
    n_steps=5,
    order=1,
)
\end{lstlisting}
Here, \texttt{time} is the total evolution time, \texttt{n\_steps} is the number of Trotter steps, and \texttt{order} is the order of the product formula.

\subsection{Efficient sampling of Slater determinants}
\label{sec:slater-sampling}

Slater determinants are an important class of fermionic states that admit efficient classical description via the orbital rotation $U$ and electronic configuration $\ket{I_\alpha, I_\beta}$ (see Section~\ref{sec:background-slater}). Finding the variationally optimal single determinant defines the Hartree-Fock method, and determinants are often used to construct initial states for quantum simulations~\cite{ortiz2001quantum,fomichev2024initial}. The classical tractability of Slater determinants makes them useful test cases for numerical simulations of quantum algorithms, as well as for benchmarking quantum computing hardware~\cite{google2020hartreefock}.

\ffsim{} includes functions for constructing and manipulating Slater determinants in the efficient representation. In particular, \ffsim{} can efficiently sample electronic configurations from Slater determinants. Sampling a Slater determinant $\ket{\text{Slater}}$ means sampling a configuration $(I_\alpha, I_\beta)$ according to the probability distribution $p((I_\alpha, I_\beta)) = \abs{\braket{I_\alpha, I_\beta}{\text{Slater}}}^2$. In the mathematics and machine learning literature, this probability distribution is known as a determinantal point process~\cite{kulesza2012determinantal}, and efficient sampling algorithms are known which can scale to thousands of orbitals. \ffsim{} implements the algorithm from Ref.~\cite{sun2023fermion}, which has time complexity $O(\eta^2 N)$ where $\eta = \max(N_\alpha, N_\beta)$.

The following code samples 1,000 independent configurations from a Slater determinant specified by an orbital rotation and the reference configuration $\ket{\set{0, 1}, \set{0, 1}}$:
\begin{lstlisting}
samples = ffsim.sample_slater(norb, ([0, 1], [0, 1]), orbital_rotation, shots=1000)
\end{lstlisting}

\subsection{General fermionic operators}
\label{sec:general-operators}

\ffsim{} includes a \texttt{FermionOperator} class for representing arbitrary fermionic operators as linear combinations of products of creation and annihilation operators, with coefficients stored in a dictionary (hash table). It is similar to the identically named class in \openfermion{}~\cite{openfermion}, but is implemented internally using Rust, making it much faster. In \ffsim{}, the \texttt{FermionOperator} class is mainly used to build model Hamiltonians, which are often subsequently converted to an array-based data structure for further processing.

\begin{figure}
    \includegraphics[width=\linewidth]{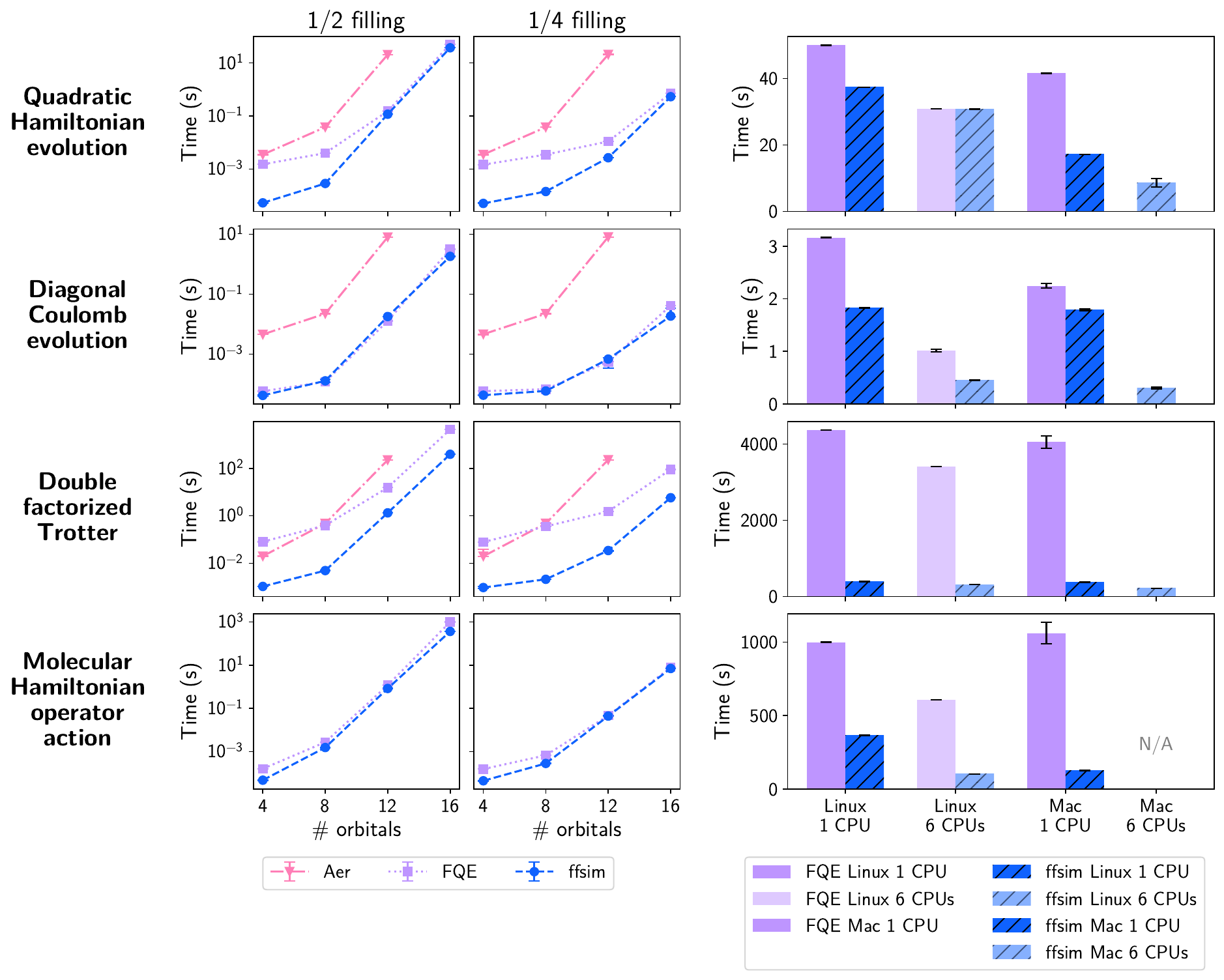}
    \caption{Performance benchmarks for state vector simulation: Time evolution by a quadratic Hamiltonian; time evolution by a diagonal Coulomb operator; Trotter simulation by a double factorized Hamiltonian (three order-0 Trotter steps of a Hamiltonian with three terms); operator action of a molecular Hamiltonian with real-valued coefficients.
    \textbf{Left}: Single-threaded wall-clock time for state vector simulation as a function of number of spatial orbitals, on a Linux x86 laptop (see main text for machine details). Data points show the median time taken to perform the operation, and error bars show the interquartile range. Data is shown for both $1/2$ filling and $1/4$ filling. At a fixed number of spatial orbitals, \ffsim{} and \fqe{} are faster at $1/4$ filling than at $1/2$ filling because the simulated symmetry sector is smaller. \aer{} stores all $2^{2N}$ amplitudes for $N$ spatial orbitals regardless of filling, so lower filling does not provide the same reduction in simulation cost.
    \textbf{Right}: Timing results for the largest instance of each state vector simulation task, 16 spatial orbitals at $1/2$ filling. Bar height shows the median time taken to perform the operation, and error bars show the interquartile range. Data is shown for both single-threaded and multi-threaded (with 6 threads) configurations, and for measurements taken from both a Linux x86 laptop and a MacBook with Apple silicon. We omit multi-threaded MacBook data for \fqe{}; see the main text for an explanation of missing data.
    }
    \label{fig:sim}
\end{figure}

\section{Comparison with other software}
\label{sec:comparison}

\subsection{Comparison with \fqe{}}
\label{sec:comparison-fqe}

\fqe{}~\cite{rubin2021fqe} is an open-source Python library that shares some functionality with \ffsim{}. While both libraries can evolve fermionic wave functions with particle number and spin~$z$ symmetry, they differ in software design, performance, and features. Table~\ref{tab:comparison-fqe} compares their interfaces and selected capabilities.

\begin{table}[!t]
    \centering
    \small
    \renewcommand{\arraystretch}{1.1}
    \begin{tabularx}{\textwidth}{@{}>{\raggedright\arraybackslash}p{0.21\textwidth} >{\raggedright\arraybackslash}X >{\raggedright\arraybackslash}X@{}}
        \toprule
        \textbf{Aspect} & \textbf{\ffsim{}} & \textbf{\fqe{}} \\
        \midrule
        Spinful state vectors & One-dimensional \numpy{} array for a fixed particle-number and spin-$z$ sector. & \texttt{Wavefunction} object containing one or more particle-number/spin-$z$ sectors; can break either symmetry while preserving the other. \\
        \addlinespace
        Software design & Functions acting on arrays; operator action via \scipy{} \texttt{LinearOperator}. & Wave function methods and functions; library-specific Hamiltonian classes. \\
        \addlinespace
        \pyscf{} interoperability & Compatible FCI addressing; uses \pyscf{} for molecular Hamiltonian action. & Different addressing requires conversion; own implementation of Hamiltonian action. \\
        \addlinespace
        Orbital rotations & Givens decomposition, also used for quadratic and double factorized Trotter evolution. & LU decomposition for basis transformations; separate Givens implementation in double factorized Trotter evolution. \\
        \addlinespace
        Quantum software interfaces & \qiskit{} circuit simulation, \texttt{Sampler}, gate decompositions, and transpiler passes. & \openfermion{} operator input and conversion to and from \textsf{Cirq} state vectors. \\
        \addlinespace
        Variational interfaces & Dedicated UCJ/LUCJ and UCCSD classes; linear method optimizer. & ADAPT-VQE with \scipy{} optimization. \\
        \addlinespace
        Other built-in tools & Compressed double factorization; efficient Slater determinant sampling. & Davidson diagonalization; Brillouin-condition solvers. \\
        \bottomrule
    \end{tabularx}
    \caption{Comparison of interfaces and selected built-in tools in \ffsim{} and \fqe{}. Performance benchmarks are presented in Section~\ref{sec:performance}.}
    \label{tab:comparison-fqe}
\end{table}

\ffsim{} uses a functional programming style, with state vectors represented directly as \numpy{} arrays and state manipulation performed with functions that accept the state vector as input. \fqe{} opts for a more object-oriented style, introducing a specialized class to represent wave functions, with state manipulation performed with either class methods or functions. In \ffsim{}'s spinful representation, each array stores a single sector of fixed particle number and spin~$z$, whereas \fqe{}'s \texttt{Wavefunction} class can store multiple sectors. \fqe{} also supports evolution that breaks either particle-number or spin~$z$ symmetry while preserving the other.

Another way that \ffsim{} differs from \fqe{} is in its use of, and integration with, the widely used chemistry library \pyscf{}~\cite{pyscf-1,pyscf-2,pyscf-3}. Because \ffsim{} uses \pyscf{}'s FCI module to assign state vector addresses, state vectors can be directly passed between the two libraries with full compatibility. On the other hand, \fqe{} implements a different addressing scheme, so data conversion is required when working with both \fqe{} and \pyscf{}. \ffsim{} also uses \pyscf{}'s FCI module to implement operator action by a molecular Hamiltonian, and we found it to be faster than \fqe{}'s implementation (see Section~\ref{sec:performance}).

The implementation of orbital rotations is a major difference between \ffsim{} and \fqe{}. While \ffsim{} uses a Givens rotation decomposition as described in Section~\ref{sec:orbital-rotation}, \fqe{} uses an approach based on the LU decomposition~\cite{malmqvist1986calculation,mitrushchenkov2007calculation}. The LU factorization costs $O(N^3)$, and applying the transformation requires $O(N)$ column updates per spin, each accumulating at most $O(N)$ single-excitation contributions per amplitude, so the state vector cost is $O(N^2 D)$. Thus, both the Givens and the LU approaches have the same asymptotic scaling.

The implementations differ in the number and organization of state vector updates. The LU decomposition can introduce a permutation of the orbitals, which must be tracked or later reversed. In \fqe{}'s quadratic Hamiltonian evolution routine, the state is rotated to the orbital basis that diagonalizes the Hamiltonian, evolved in that basis, and then rotated back, undoing the permutation in the process. In contrast, \ffsim{} forms the $N \times N$ matrix $U = \exp(-iMt)$ and applies it as a single orbital rotation, which reduces the number of full basis transformations from two to one.

For double factorized Trotter evolution, \fqe{}'s \texttt{double\_factor\_trotter\_evolution} function uses a separate Givens rotation implementation rather than its optimized LU code. Each rotation goes through a general evolution routine that copies the state vector and updates it in several passes. \ffsim{} applies each rotation directly in place using a dedicated kernel. Both methods scale as $O(N^2 D)$, but \ffsim{} reduces copying and other overhead. These implementation differences help explain the timing results in Section~\ref{sec:performance}.

Finally, while both libraries share the core feature of state vector simulation, their other features have naturally diverged. \ffsim{} provides \qiskit{} circuit simulation and compilation tools (Section~\ref{sec:qiskit}), while \fqe{} accepts \openfermion{} operators and provides conversion to and from \textsf{Cirq} state vectors. Section~\ref{sec:additional-features} describes additional \ffsim{} tools, including dedicated UCJ/LUCJ classes, compressed double factorization, and efficient sampling of Slater determinants without constructing the full state vector. Examples of additional \fqe{} tools include ADAPT-VQE, Davidson diagonalization, and solvers for the two-particle Brillouin condition. We refer the reader to the documentation of both libraries for more details.

\begin{figure}
    \includegraphics[width=\linewidth]{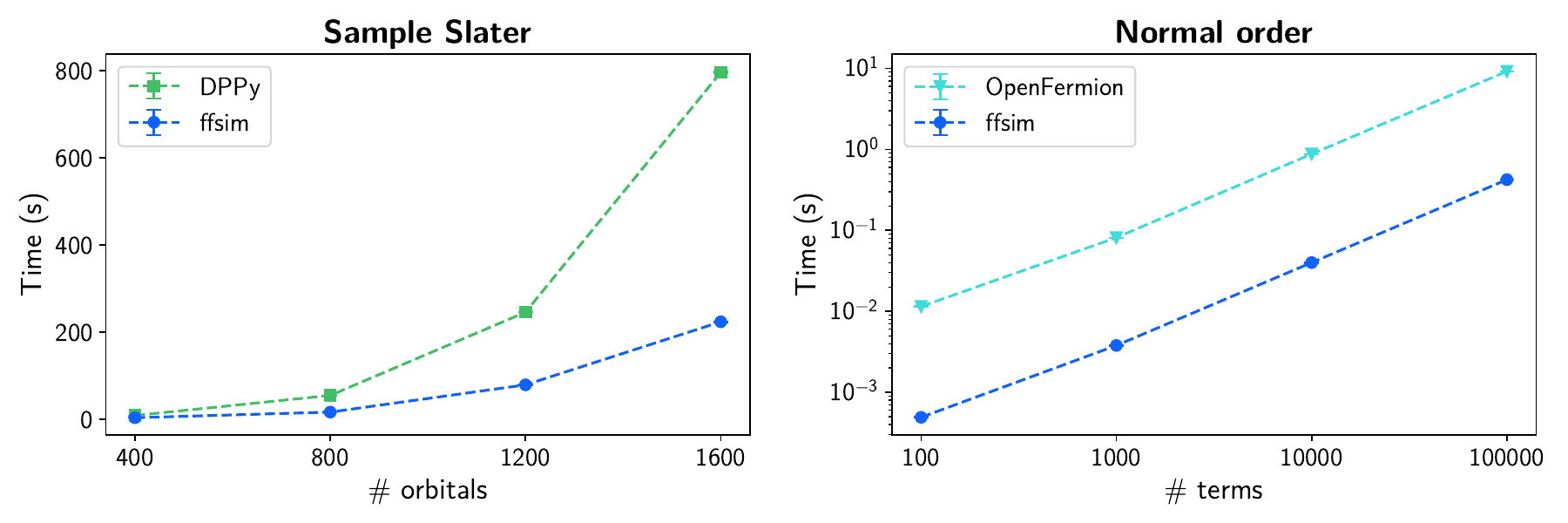}
    \caption{
    Performance comparison for additional features, as measured on a Linux x86 laptop (see main text for machine details). Data points show the median time taken to perform the operation, and error bars show the interquartile range.
    \textbf{Left}: Time taken to sample 1000 configurations from a spinless Slater determinant at $1/4$ filling.
    \textbf{Right}: Time taken to normal order a random fermionic operator with 100 spatial orbitals and varying numbers of terms.
    }
    \label{fig:extra}
\end{figure}

\subsection{Performance benchmarks}
\label{sec:performance}

We compared \ffsim{}'s performance in state vector simulation against \fqe{} and \aer{}~\cite{qiskit,qiskit_aer}. \aer{} is a general-purpose quantum circuit simulator, so to compare against it, we compiled the fermionic gates to quantum circuits in the Jordan-Wigner representation using the same strategies illustrated in Figure~\ref{fig:qiskit-lucj-circuit}. We also compared \ffsim{}'s performance in sampling from Slater determinants against \dppy{}~\cite{dppy}, and its performance in normal ordering fermionic operators against \openfermion{}~\cite{openfermion}. We collected the timing measurements using \asv{}~\cite{asv}. For each benchmark, the number of measurements to collect was automatically determined by \asv{} using its default settings, and the required objects (for example, Hamiltonian and state vector) were generated randomly (see the source code, linked to at the end of this paper, for details). Data for Linux on x86 were collected on a Lenovo ThinkPad X13 (Gen 1) with AMD Ryzen 7 Pro 4750U and 32 GB RAM. Data for MacBook with Apple Silicon were collected on a 16-inch MacBook Pro (2021) with Apple M1 Max and 64 GB RAM. We compiled \ffsim{} with \texttt{-C target-cpu=native}, \pyscf{} with \texttt{-DBUILD\_MARCH\_NATIVE=ON}, and \fqe{} with its default options (which includes \texttt{-march=native} on Linux). We used the following software library versions: \ffsim{}~\texttt{0.0.79}, \pyscf{}~\texttt{2.12.1},~\fqe{} \texttt{0.3.0}, \dppy{}~\texttt{0.3.3}, \openfermion{}~\texttt{1.7.1}.

Figure~\ref{fig:sim} shows timing results for state vector simulation. The left side of the figure shows single-threaded run times of \ffsim{}, \fqe{}, and \aer{} on state vector simulation tasks as a function of number of spatial orbitals. The right side of the figure restricts to the largest instance of each task, 16 spatial orbitals at $1/2$ filling, and shows both single- and multi-threaded timing results for \ffsim{} and \fqe{}, as well as data for both Linux on x86 and a MacBook with Apple Silicon. The documented method for enabling multi-threading (via the \texttt{OMP\_NUM\_THREADS} environment variable) had no apparent effect for \fqe{} on the MacBook, so we omit multi-threaded data for \fqe{} on the MacBook. We found the same to be true of \pyscf{}, which \ffsim{} uses for molecular Hamiltonian action, so we also omit multi-threaded data for \ffsim{} for that task on the MacBook. Figure~\ref{fig:extra} shows run time comparisons for sampling from Slater determinants and normal ordering fermionic operators, with data collected on the Linux x86 laptop.

Across all state vector simulation benchmarks, \ffsim{} outperforms \fqe{} for the largest system size considered. On the single-threaded MacBook benchmarks, \ffsim{} achieved speedups of $2.4\times$, $1.2\times$, and $8.4\times$ for quadratic Hamiltonian evolution, diagonal Coulomb evolution, and Molecular Hamiltonian operator action, respectively. When allowed to use multiple threads, the speedups increased to $4.8\times$ and $7.5\times$ for quadratic Hamiltonian evolution and diagonal Coulomb evolution. For double factorized Trotter simulation, \ffsim{} achieves more dramatic speedups of up to $18\times$. Section~\ref{sec:comparison-fqe} discusses implementation differences that contribute to these speedups, including \fqe{}'s use of a separate Givens rotation implementation for this task. The orbital rotation algorithms have the same asymptotic scaling, so these timings reflect differences in implementation and constant factors. For the benchmarks where \aer{} is applicable, it lags far behind both libraries and is unable to simulate the largest system of 16 spatial orbitals, which maps to 32 qubits. A general-purpose state vector simulator like \aer{} would require 64 GiB to store the state vector for this system using double precision, while \ffsim{} and \fqe{} need only 2.5 GiB at $1/2$ filling and 51 MiB at $1/4$ filling. For sampling from Slater determinants and normal ordering fermionic operators, \ffsim{} achieves a $3.5\times$ speedup over \dppy{} and $20\times$ speedup over \openfermion{}.

\begin{figure}
    \includegraphics[width=\linewidth]{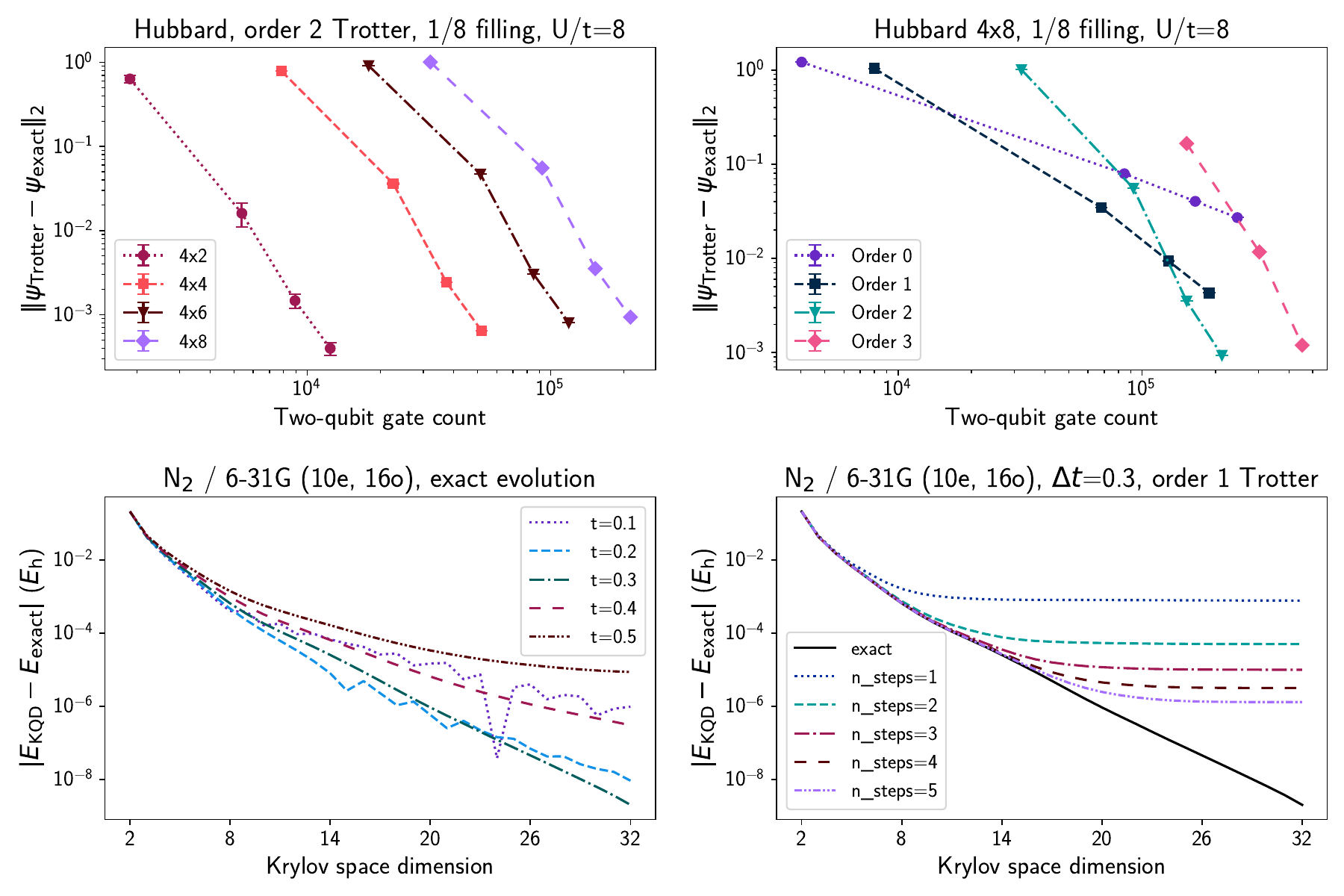}
    \caption{Applications.
    \textbf{Top left}: Trotter error of the order-2 product formula as a function of two-qubit gate count for the 2D Hubbard model at $1/8$ filling for different lattice sizes. Data points show the average error over 5 random vectors sampled from the uniform distribution over the unit sphere, and error bars indicate $\pm 1$ standard deviation. Note that the order-2 product formula is commonly referred to as the fourth-order formula in the literature (see Section~\ref{sec:hamiltonian-time-evolution}).
    \textbf{Top right}: Trotter error as a function of two-qubit gate count for the $4 \times 8$ Hubbard model at $1/8$ filling, for different product formula orders.
    \textbf{Bottom left}: Absolute ground state energy error of Krylov Quantum Diagonalization (KQD) applied to N\textsubscript{2}/6-31G (10e,~16o) as a function of Krylov space dimension, for different choices of time step.
    \textbf{Bottom right}: Absolute ground state energy error of KQD with Trotterized time evolution as a function of Krylov space dimension. The time step for the time evolution operator is fixed to $\Delta t = 0.3$, but the different lines show data for different numbers of Trotter steps used to approximate the time evolution operator using the order-1 product formula. Higher numbers of Trotter steps yield better approximations, and the solid black line shows data for KQD with exact time evolution. Note that the order-1 product formula is commonly referred to as the second-order formula in the literature (see Section~\ref{sec:hamiltonian-time-evolution}).
    }
    \label{fig:applications}
\end{figure}

\section{Applications}
\label{sec:applications}

\subsection{Sample-based algorithms}

A number of recent experiments on quantum computing hardware have used algorithms that do not require quantum estimation of expectation values, a challenging subroutine for current noisy quantum processors~\cite{robledomoreno2025chemistry,yu2025skqd,liepuoniute2025quantumcentric,kaliakin2025implicit,wang2025samplebased,piccinelli2026quantum}. Instead, these sample-based algorithms post-process sampled bitstrings without needing to extract expectation values from them~\cite{kanno2023qsci,mikkelsen2025qsci}. \ffsim{} is well-suited for studying these algorithms because its state vector representation can readily be sampled, as squaring the absolute values of the entries yields the full set of probabilities.

Previous works have used \ffsim{} to simulate sample-based algorithms for realistic chemical systems beyond the reach of general-purpose state vector simulators. In Ref.~\cite{robledomoreno2025chemistry}, \ffsim{} was used to study the expressiveness of the LUCJ ansatz for representing the ground state of the iron-sulfur cluster [2Fe-2S] in an active space of 30 electrons in 20 orbitals, which maps to 40 qubits under the Jordan-Wigner transformation. In Ref.~\cite{lin2025improved}, \ffsim{} was used to study different ways of initializing the parameters of the LUCJ ansatz applied to N\textsubscript{2} in an active space of 10 electrons in 26 orbitals, which maps to 52 qubits. A general-purpose quantum circuit simulator would require 64 PiB (about $6.7 \times 10^{7}$ GiB) to store the state vector of 52 qubits using double precision, an amount that exceeds the total memory of even the world's most powerful supercomputers. On the other hand, \ffsim{} needs only 64.5 GiB for this system, allowing it to be simulated on a single workstation. The interested reader can see the references for more details.

\subsection{Estimating Trotter error}
\label{sec:trotter_error}

Trotter-Suzuki product formulas are a common approach to approximating Hamiltonian time evolution on a quantum computer (see Section~\ref{sec:hamiltonian-time-evolution}). The difference between the approximate and exact evolutions is referred to as the Trotter error, and it can be made arbitrarily small by increasing the number of time steps or the order of the product formula. When implementing Trotterized time evolution on a quantum computer with a fixed budget on circuit operations, estimates of Trotter error inform both the size of the system to target as well as the choice of product formula order.

Upper bounds on the Trotter error can be obtained from nested commutators of the Hamiltonian terms~\cite{childs2021theory}. For fermionic Hamiltonians, Ref.~\cite{su2021nearly} derives bounds that exploit particle-number conservation and the structure of the interactions, while Ref.~\cite{schubert2023trotter} evaluates nested commutators for the Fermi-Hubbard model on one- and two-dimensional lattices to obtain explicit higher-order bounds. Such bounds can still overestimate the error for specific systems~\cite{childs2021theory,schubert2023trotter}. Direct numerical computation of the Trotter error for small representative systems provides a complementary approach that can give insights on the error for larger systems~\cite{childs2018toward}.

Here, we use \ffsim{} to compute the Trotter error for simulation of the two-dimensional Hubbard model. We quantify this error as the Euclidean distance $\lVert{\psi_\text{Trotter} - \psi_\text{exact}}\rVert_2$ between the state resulting from the Trotterized evolution and the state resulting from exact time evolution. Given an observable of interest, this distance bounds the difference between the corresponding expectation values. We apply the evolutions to five random state vectors sampled from the uniform distribution over the complex unit sphere, and report the average error and standard deviation.

The 2D Hubbard model Hamiltonian is
\begin{align}
    H = -t \sum_{\langle pq \rangle, \sigma}
    \parens*{a^\dagger_{p \sigma} a_{q \sigma} + a^\dagger_{q \sigma} a_{p \sigma}}
    + U \sum_p n_{p\alpha} n_{p\beta},
\end{align}
where the indices $\langle pq \rangle$ range over the edges of a two-dimensional square lattice, and in this study we set $t=1$ and $U=8$. This Hamiltonian is a diagonal Coulomb Hamiltonian and we Trotterize it as the sum of two terms, as described in Section~\ref{sec:hamiltonian-time-evolution}. We fix the total evolution time to $T=1$. While the Hubbard model is typically studied near $1/2$ filling~\cite{hubbard1963electron,zheng2017stripe}, this regime is challenging for \textsf{ffsim} because of the large Hilbert space dimension. Instead, we study the model at $1/8$ filling. We fix the $x$ dimension to $4$ and let the $y$ dimension range from $2$ to $8$, and use periodic boundary conditions in the $x$ direction. The largest model we simulate is a $4 \times 8$ Hubbard lattice with 32 sites and eight electrons, four of each spin ($N_\alpha = N_\beta = 4$), which maps to 64 qubits under the Jordan-Wigner transformation. The low filling makes this example tractable: the state vector dimension is $D = \binom{32}{4}^2 = 1{,}293{,}121{,}600$, so using complex double precision to store the amplitudes requires $16D$ bytes, or 19.3 GiB.

For each choice of system size, product formula order, and number of Trotter steps, we compiled a quantum circuit implementing the Trotterized evolution (assuming no restrictions on qubit connectivity) and counted the number of two-qubit gates. We used this two-qubit gate count to represent the quantum cost of the simulation, enabling a fair comparison between different choices of parameters. The compilation is deterministic, with orbital rotations and diagonal Coulomb evolutions implemented as illustrated in Figure~\ref{fig:qiskit-lucj-circuit}.

The top left panel of Figure~\ref{fig:applications} plots the Trotter error as a function of two-qubit gate count for different lattice sizes, with the product formula fixed to order 2. As the gate budget is increased, more Trotter steps (each with a smaller step size) can be performed, and the error decreases. Larger lattice sizes require more gates to achieve a given error. The top right panel of the figure plots the Trotter error for different product formula orders, with the lattice size fixed to $4 \times 8$. While higher-order formulas generally have better asymptotic scalings, the optimal choice of product formula order for a given problem depends on the exact gate budget or the desired error tolerance.

\subsection{Krylov quantum diagonalization}

Quantum subspace methods are a promising choice for approximating Hamiltonian eigenstates and energies on quantum computers. These methods compute the projection of the Hamiltonian onto the subspace spanned by a small number of vectors, yielding a generalized eigenvalue problem that can be efficiently solved on a classical computer. Here, we show how \ffsim{} can be used to study Krylov quantum diagonalization (KQD)~\cite{parrish2019quantumfilterdiagonalization,stair2020multireference,yoshioka2025krylov}. In KQD, the vectors spanning the subspace are obtained by applying powers of a time evolution operator to a reference vector. The resulting vectors are called Krylov vectors and the subspace is called a Krylov space. Explicitly, the Krylov vectors are $\set{\exp(-i H k \Delta t) \ket{\Psi}}_{k = 0}^{D - 1}$, where $\ket{\Psi}$ is the reference vector (here taken to be the Hartree-Fock state), $D$ is the Krylov space dimension, and $\Delta t$ is the time step. We study KQD applied to approximating the ground state energy of N\textsubscript{2} in a (10e,~16o) active space (32 qubits) derived from the 6-31G basis set.

An important hyperparameter in KQD is the time step $\Delta t$ to use for the time evolution operator. While theory suggests a value of $\pi$ divided by the spectral norm of the Hamiltonian~\cite{parrish2019quantumfilterdiagonalization,epperly2022theory}, a good estimate of this value may be unavailable, and in any case, the optimal choice depends on the initial state. Consequently, experimental implementations of KQD and related algorithms often choose the time step based on heuristics informed by numerical simulation~\cite{yoshioka2025krylov,yu2025skqd,piccinelli2026quantum}. The bottom left panel of Figure~\ref{fig:applications} plots the energy error of KQD as a function of the Krylov space dimension, for various choices of time step. When the time step is too small, KQD suffers from numerical instability because the Krylov vectors are very close to each other, causing the generalized eigenvalue problem to be nearly singular. As a result, the error may fail to decrease monotonically with increased Krylov space dimension. On the other hand, when the time step is too large, the energy converges slowly with increased dimension.

Digital quantum computers cannot perform exact time evolution, and instead must use approximate methods such as Trotter-Suzuki product formulas. While the error can be made arbitrarily small with increased circuit depth, KQD can perform well even when the exact time evolution is only roughly approximated using short-depth circuits, which enables demonstrations on existing noisy quantum processors and saves resources on future fault-tolerant quantum computers~\cite{epperly2022theory,kirby2024analysis}. The bottom right panel of Figure~\ref{fig:applications} shows the energy error of KQD with the time evolution operator approximated using a Trotter-Suzuki product formula of fixed order. Different levels of approximation are shown, with increased number of Trotter steps indicating higher accuracy (and higher circuit depth). The molecular Hamiltonian is Trotterized in the double factorized representation as described in Section~\ref{sec:hamiltonian-time-evolution}.

\section{Limitations}
\label{sec:limitations}

Exploiting particle number and spin $z$ symmetries extends the range of tractable simulations but does not remove the exponential scaling of exact state vector simulation at fixed filling. For $N$ spatial orbitals, \ffsim{} stores $D = \binom{N}{N_\alpha}\binom{N}{N_\beta}$ complex amplitudes. At half filling with $N_\alpha = N_\beta = N/2$, Stirling's approximation gives $D \sim 2 \cdot 4^N/(\pi N)$, so the reduction relative to the full $4^N$-dimensional Hilbert space is only a factor proportional to $N$. At lower fixed filling fractions, $D$ grows more slowly but remains exponential in $N$. If instead the particle numbers are held fixed as $N$ increases, the growth is polynomial, $D = O(N^{N_\alpha+N_\beta})$, although the degree can still make simulations impractical.

Storing one state vector in complex double precision requires $16D$ bytes. The 32-orbital example in Section~\ref{sec:trotter_error} requires 19.3 GiB at $1/8$ filling, whereas the same number of orbitals at half filling with equal spin populations would require about 5.0 EiB. These figures exclude temporary arrays, cached address maps, and other overhead. Runtime is also a constraint: a dense orbital rotation requires $O(N^2D)$ state vector work at fixed filling (see Section~\ref{sec:orbital-rotation}), and this cost accumulates with circuit depth or the number of Trotter steps. Faster kernels and parallel execution can improve practical performance but do not change this asymptotic scaling.

The spinful state vector representation is restricted to a single sector of fixed particle number and spin~$z$. Evolution that couples different sectors, such as particle-number-changing or spin-flip interactions, cannot be simulated directly within that representation. The \qiskit{} simulation interface likewise requires the circuit structure described in Section~\ref{sec:qiskit}, including Hamming weight-preserving evolution and measurements only at the end. The compact Slater determinant routines in Section~\ref{sec:slater-sampling} avoid full state vector storage for that special class of states, but do not provide a compact representation of general correlated states produced by interacting circuits.

\section{Conclusion}
\label{sec:conclusion}

We have presented \ffsim{}, an open-source software library for fast simulation of fermionic quantum circuits. By exploiting conservation of particle number and the $z$~component of spin, \ffsim{} achieves substantial efficiency gains over general-purpose quantum circuit simulators and outperforms \fqe{}, a library with similar functionality. In addition to state vector simulation, \ffsim{} offers features including variational ansatzes, Hamiltonian time evolution, efficient sampling of Slater determinants, integration with \qiskit{} and \pyscf{}, and comprehensive documentation. We demonstrated the capabilities of \ffsim{} on scientific applications involving circuits of up to 64 qubits.

Looking ahead, we believe GPU acceleration to be a promising direction for further performance improvements, as the application of fermionic gates to a state vector can be parallelized over the state vector entries. Such acceleration may provide significant speedups and further extend the range of classically tractable systems.

As quantum computing hardware continues to mature, we hope \ffsim{} will serve as a useful tool for the quantum computing community as they develop and test algorithms for these emerging systems.

\section*{Code availability}

The source code for this work is available at the following URLs:
\begin{itemize}
    \item \ffsim{}: \url{https://github.com/Qiskit/ffsim}
    \item Performance benchmarks: \url{https://github.com/kevinsung/ffsim-benchmark}
    \item Applications: \url{https://github.com/kevinsung/ffsim-numerics}
\end{itemize}

\section*{Author contributions}

KJS designed \ffsim{} and serves as its lead developer. All authors contributed to the development of \ffsim{} by discussing features, writing code, or writing documentation. Generative AI was used to write some code and parts of the manuscript.

\acknowledgments{
KJS thanks Will Kirby for helpful discussions about Krylov quantum diagonalization.
BA acknowledges support from the Swiss National Science Foundation under grant nos. \texttt{P500PT\_203168} and \texttt{P5R5PT\_225346}, as well as the US Department of Energy, Office of Science, Basic Energy Sciences, Early Career Award no. \texttt{DE-SC0022716}.
}

\bibliographystyle{quantum}
\bibliography{main}

\end{document}